\documentclass[smallabstract,smallcaptions]{dccpaper}

\usepackage{amsmath}
\usepackage{amssymb}
\usepackage{color}
\usepackage{url}
\usepackage{amsfonts}
\usepackage{graphicx}
\usepackage{hyperref}
\usepackage{booktabs}
\usepackage{colortbl}
\usepackage{underscore}
\usepackage{placeins}
\usepackage{array}
\usepackage{multirow}
\usepackage{enumitem}

\definecolor{lightgray}{gray}{0.9}

\newlength{\figurewidth}
\newlength{\smallfigurewidth}
\newcommand{\wtilde}[1]{\stackrel{\sim}{\smash{#1}\rule{0pt}{1.1ex}}}
\newcommand{\wstilde}[1]{\stackrel{\sim}{\smash{#1}\rule{0pt}{0.8ex}}}

\begin{document}

\title
{\large
\textbf{Learned Compression for Compressed Learning}
}

\author{%
Dan Jacobellis and Neeraja J. Yadwadkar\\[0.5em]
{\small\begin{minipage}{\linewidth}\begin{center}
\begin{tabular}{ccc}
University of Texas at Austin \\
Austin, TX, 78712, USA\\
\url{danjacobellis@utexas.edu} \\
\url{neeraja@austin.utexas.edu} 
\end{tabular}
\end{center}\end{minipage}}
}

\maketitle
\thispagestyle{empty}

\begin{abstract}

Modern sensors produce increasingly rich streams of high-resolution data.
Due to resource constraints, machine learning systems discard the vast majority of this information via resolution reduction.
Compressed-domain learning allows models to operate on compact latent representations, allowing higher effective resolution for the same budget. 
However, existing compression systems are not ideal for compressed learning.
Linear transform coding and end-to-end learned compression systems reduce bitrate, but do not uniformly reduce dimensionality; thus, they do not meaningfully increase efficiency.
Generative autoencoders reduce dimensionality, but their adversarial or perceptual objectives lead to significant information loss.
To address these limitations, we introduce WaLLoC (\textbf{Wa}velet \textbf{L}earned \textbf{Lo}ssy \textbf{C}ompression), a neural codec architecture that combines linear transform coding with nonlinear dimensionality-reducing autoencoders.
WaLLoC sandwiches a shallow, asymmetric autoencoder and entropy bottleneck between an invertible wavelet packet transform.
Across several key metrics, WaLLoC outperforms the autoencoders used in state-of-the-art latent diffusion models. 
WaLLoC does not require perceptual or adversarial losses to represent high-frequency detail, providing compatibility with modalities beyond RGB images and stereo audio.
WaLLoC's encoder consists almost entirely of linear operations, making it exceptionally efficient and suitable for mobile computing, remote sensing, and learning directly from compressed data.
We demonstrate WaLLoC's capability for compressed-domain learning across several tasks, including image classification, colorization, document understanding, and music source separation. Our code, experiments, and pre-trained audio and image codecs are available at \url{https://ut-sysml.org/walloc/}.

\end{abstract}

\section{Introduction}
\vspace{-3mm}

\begin{figure}[t]
    \centering
    \includegraphics[width=1\linewidth]{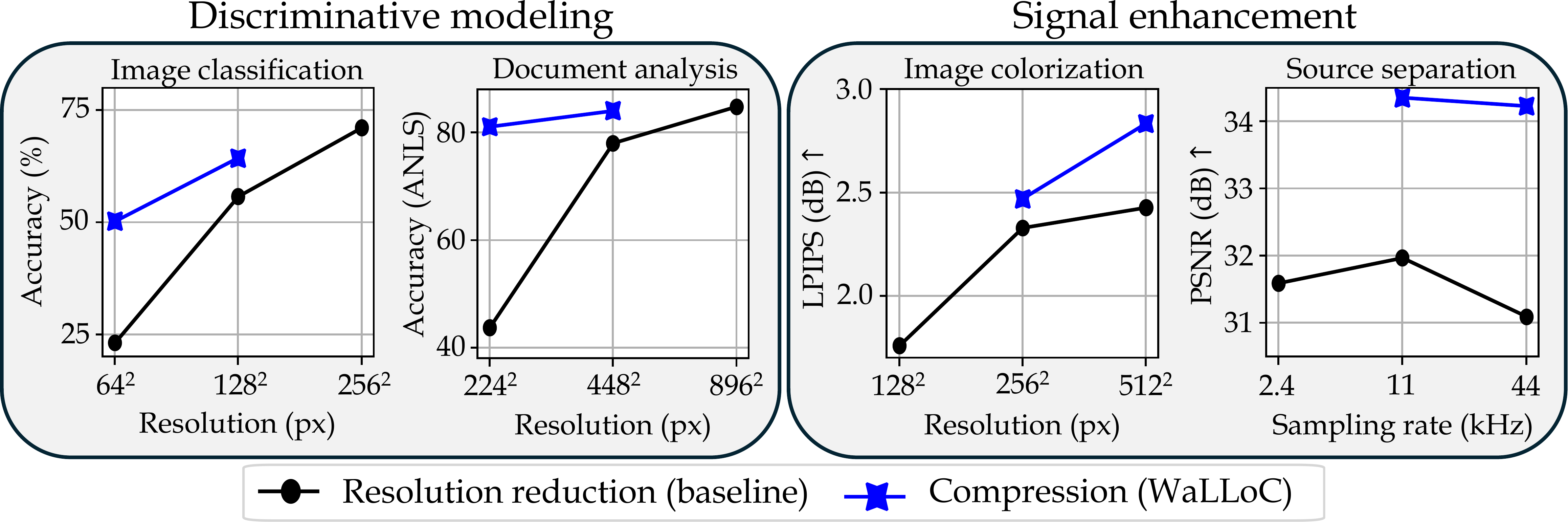}
    \vspace{-9mm}
    \caption{In discriminative models (left), resolution reduction increases training and inference efficiency, but significantly degrades accuracy. Replacing resolution reduction with WaLLoC leads to significantly higher accuracy, while providing the same degree of acceleration. For signal enhancement (right), WaLLoC provides better quality when scaling to high resolutions compared to directly operating on image pixels or audio samples. 
    }
    \label{fig:resolution}
\end{figure}

In the last decade, deep neural networks (DNNs) have rapidly evolved from simple classifiers~\cite{krizhevsky2012imagenet,hershey2017cnn} to domain-specific and multi-modal foundation models~\cite{archit2023segment, beyer2024paligemma}.
With this shift, models are increasingly able to make use of minute and high-frequency signal details.
For example, when increasing the resolution of PaliGemma from $224^2$ to $896^2$ pixels (Figure \ref{fig:resolution}), its ability to analyze documents increases from 44\% to 85\% ANLS~\cite{beyer2024paligemma}. However, operating at this increased resolution requires significantly more GPU memory (21 vs 8 GB) and leads to 4$\times$ higher latency.

Compressed-domain learning~\cite{ehrlich2019deep,park2023storage,rombach2022high} has been proposed to improve the trade-off between model accuracy and compute needs. 
In this paradigm, the model operates on low-dimensional (lossy) compressed data, thereby enabling dramatic reductions in compute cost and inference latency while maintaining model accuracy.
However, existing lossy compression methods, coming from three main categories, are not ideal for compressed-domain learning. 
(a) Linear transform coding methods (e.g., JPEG, MP3) reduce bitrate via energy-compacting time-frequency transforms, but do not meaningfully reduce dimensionality or increase efficiency of downstream models.
(b) End-to-end learned codecs~\cite{balle2017end} achieve better rate-distortion performance and modestly reduce dimension via nonlinear autoencoders, but high encoding overhead negates the benefits of compressed learning.
(c) Generative autoencoders~\cite{rombach2022high,evans2024stable} significantly reduce dimension, but do so by synthesizing rather than preserving details---leading to poor performance in discriminative tasks~\cite{goldblum2024battle}.

\begin{figure}[t]
    \centering
    \includegraphics[width=0.9\linewidth]{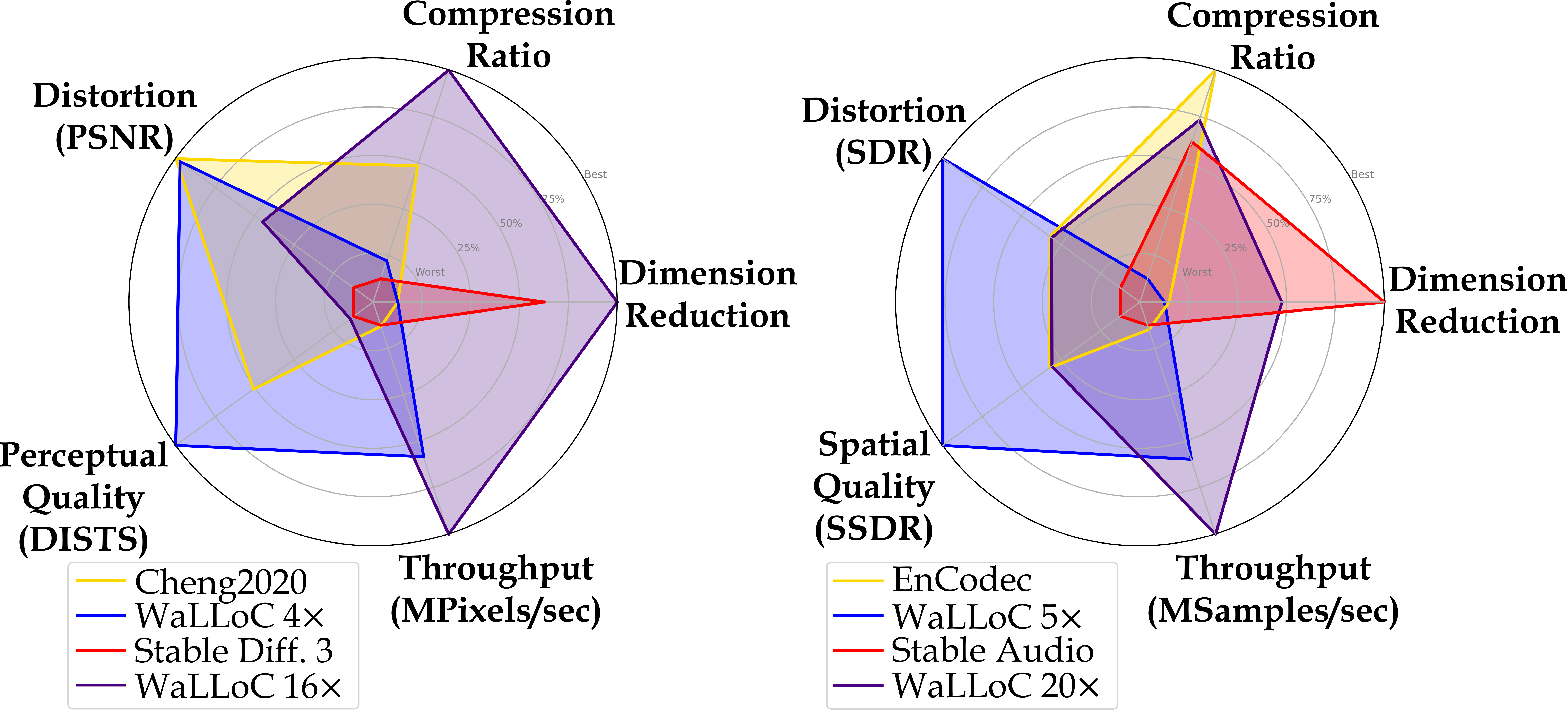}
    \caption{Comparison of our proposed method (WaLLoC) with other autoencoder designs for RGB Images (Cheng2020~\cite{cheng2020learned}, Stable Diffusion 3~\cite{esser2024scaling}) and stereo audio (EnCodec~\cite{defossez2022high}, Stable Audio~\cite{evans2024stable}). Additional metrics are reported in Tables \ref{tab:RGB} and \ref{tab:stereo}.}
    \label{fig:radar}
\end{figure}

In this work, we introduce WaLLoC (Wavelet Learned Lossy Compression), an architecture for learned compression that simultaneously satisfies three key requirements of compressed-domain learning:
\vspace{-2mm}
\begin{enumerate}[leftmargin=*]
    \item \textbf{Computationally efficient encoding} to reduce overhead in compressed-domain learning and support resource constrained mobile and remote sensors. 
    WaLLoC uses the computationally cheap and invertible
    wavelet packet transform~\cite{mallat2008} to expose signal redundancies prior to autoencoding. This allows us to replace the encoding DNN with a single linear layer ($<$100k parameters) without significant loss in quality. As shown in Figure~\ref{fig:radar}, WaLLoC incurs less than five percent of the encoding cost compared to other neural codecs.
\vspace{-2mm}
    \item \textbf{High compression ratio} for storage and transmission efficiency.
    Lossy codecs typically achieve high compression by combining quantization and entropy coding. However, naive quantization of autoencoder latents leads to unpredictable and unbounded distortion.
    Instead, we apply additive noise during training as an entropy bottleneck~\cite{balle2017end}, leading to quantization-resilient latents.
    When combined with entropy coding, WaLLoC achieves nearly 6$\times$ higher compression ratio compared to the VAE used in Stable Diffusion 3~\cite{esser2024scaling}, despite offering a higher degree of dimensionality reduction and similar quality (Figure \ref{fig:radar}, Table \ref{tab:RGB}).
\vspace{-2mm}    
    \item \textbf{Dimensionality reduction} to accelerate compressed-domain modeling.
    WaLLoC's encoder projects high-dimensional signal patches to low-dimensional latent representations, providing a reduction of up to 20$\times$.
    This allows WaLLoC to be used as a drop-in replacement for resolution reduction while providing superior detail preservation and downstream accuracy.
\end{enumerate}

\noindent Our main contributions are as follows:
\vspace{-2mm}
\begin{itemize}[leftmargin=*]
    \item We evaluate the trade-offs between three existing approaches to lossy compression---(1) linear transform coding, (2) end-to-end learned compression, and (3) generative autoencoders. We identify key limitations of each when used as a replacement for resolution reduction in machine learning models.
    \vspace{-2mm}
    \item We introduce WaLLoC, a modality-agnostic lossy compression framework that simultaneously provides (1) efficient encoding, (2) favorable rate-distortion trade-off, and (3) uniform dimensionality reduction.
    \vspace{-2mm}
    \item Using our proposed framework, we build RGB image and stereo audio codecs that outperform other autoencoder designs across several key metrics~(Figure~\ref{fig:radar}).
    We evaluate WaLLoC's efficacy for accelerating various machine learning models via compressed domain operation. Across each of the four tasks
    ---image classification, colorization, document understanding, and music source separation---
    WaLLoC outperforms resolution reduction by a wide margin~(Figure~\ref{fig:resolution}).
\end{itemize}


\vspace{-5mm}
\section{Background: Compressed-Domain Learning}
\vspace{-2mm}

Methods for compressed-domain learning can be grouped based on the type of compression (1) linear transform coding~\cite{ehrlich2019deep}, (2) end-to-end learned compression~\cite{park2023storage, ascenso2023jpeg}, and (3) and generative autoencoders~\cite{rombach2022high,evans2024stable}. 

\vspace{-5mm}
\paragraph{Linear transform coding.}
Conventional lossy compression standards---such as JPEG and MP3~\cite{mallat2008}
---are based on linear transform coding (LTC). Linear and invertible transforms like the discrete cosine transform (DCT) or discrete wavelet transform (DWT) eliminate redundancies while concentrating signal energy into fewer coefficients nearly optimally and remaining computationally efficient. Quantization allocates bits to each frequency band according to perceptual models, leading to high compression ratios with minimal perceived distortion. LTC is often combined with resolution reduction (e.g. chroma downsampling in JPEG), but does not provide consistent or uniform dimensionality reduction. LTC can improve downstream learning ~\cite{ehrlich2019deep} but does not address the computational issues of scaling DNNs to high resolution.

\vspace{-5mm}
\paragraph{End-to-end learned compression.}
Nonlinear autoencoders that are jointly optimized for both rate and distortion~\cite{balle2017end} achieve higher compression ratios than LTC, but require more computation~\cite{minnen2023advancing} and offer limited dimensionality reduction---typically 4$\times$~\cite{cheng2020learned}.
Efficient decoding, and machine vision without decoding have been explored ~\cite{yang2023computationally,ascenso2023jpeg}, but encoding overhead remains significant.

\vspace{-5mm}
\paragraph{Generative autoencoders.}
Compressed-domain learning underpins recent breakthroughs in high-resolution 
diffusion~\cite{rombach2022high}, masked autoencoding~\cite{chang2022maskgit}, and autoregressive~\cite{copet2024simple}
generative models~\cite{rombach2022high}. These applications use a low-resolution generative model paired with a generative, adversarial, and dimensionality-reducing autoencoder (GADR-AE)---which we define as any autoencoder offering $> 4\times$ dimensionality reduction (DR) and trained using adversarial and perceptual losses~\cite{esser2021taming}. GADR-AEs produce low-dimensional latent representations that are up to 64 times smaller than the original input~\cite{evans2024stable}. However, they lose significant detail in the process, so adversarial and perceptual objectives are employed to re-synthesize details in the decoder~\cite{rombach2022high}. Existing GADR-AEs are computationally cheap compared to the generative models they enable, but expensive compared to discriminative models.
For example, compared to the widely used EfficientNet model~\cite{tan2019efficientnet}, the encoder used in Stable Diffusion's VAE has $>6\times$ more parameters (34.3M vs 5.3M) and requires $>400\times$ more GFLOPs (163 vs 0.39)~\cite{rombach2022high}.

\vspace{-2mm}
\section{Proposed Method: Design and Implementation}
\vspace{-2mm}

\begin{figure}[t]
    \centering
    \includegraphics[width=1\linewidth]{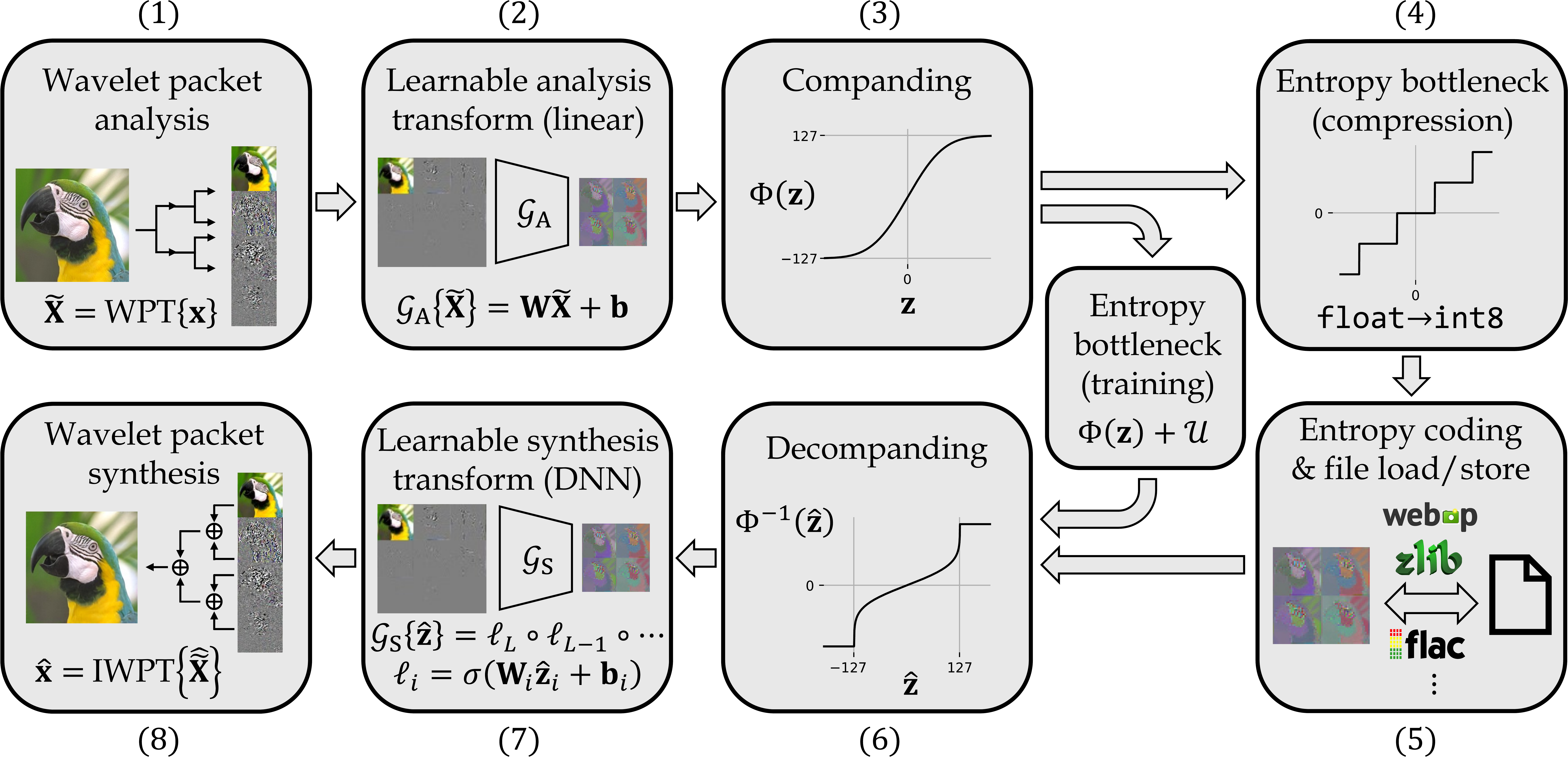}
    \vspace{-9mm}
    \caption{WaLLoC's encode-decode pipeline. The entropy bottleneck and entropy coding steps are only required to achieve high compression ratios for storage and transmission. For compressed-domain learning where dimensionality reduction is the primary goal, these steps can be skipped to reduce overhead and completely eliminate CPU-GPU transfers.}
    \label{fig:walloc}
\end{figure}

\begin{figure}
    \centering
    \includegraphics[width=.9\linewidth]{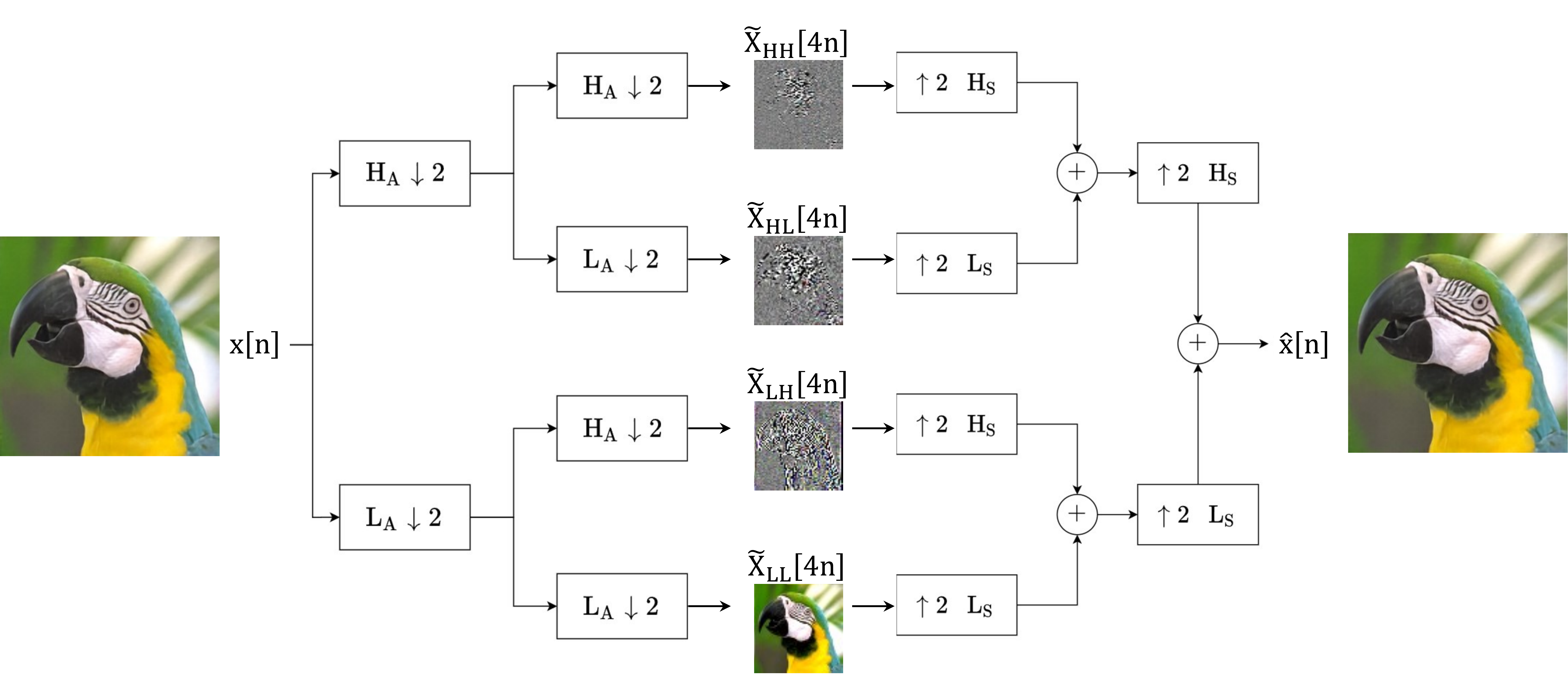}
    \vspace{-6mm}
    \caption{Example of forward and inverse WPT with $J=2$ levels.
    Each level applies filters $\text{L}_{\text{A}}$ and $\text{H}_{\text{A}}$ independently to each of the signal channels, followed by downsampling by a factor of two $\left(\downarrow 2\right)$. An inverse level consists of upsampling $\left(\uparrow 2\right)$ followed by $\text{L}_{\text{S}}$ and $\text{H}_{\text{S}}$, then summing the two channels. The full WPT $\wtilde{\textbf{X}}$ of consists of $J$ levels.}
    \label{fig:wpt}
\end{figure}

WaLLoC's design aims at achieving three goals: computationally efficient encoding, high compression ratio, and uniform dimensionality reduction. 
We note several key insights that allow us to address the limitations of previous designs that stand in the way of achieving these goals. Each of these goals, limitations, and insights motivate the core design components of WaLLoC, shown in Figure~\ref{fig:walloc}.

\vspace{-3mm}
\subsection{Achieving computationally efficient encoding.}
\vspace{-2mm}
Two main barriers stand in the way of efficient encoding.
(a) poor scaling of autoencoder performance with resolution, and (b) difficulty in preserving quality with lightweight encoders.


(a) Resolution scaling. In existing autoencoder designs ~\cite{balle2017end,cheng2020learned,rombach2022high, defossez2022high, evans2024stable}, a hierarchy of DNN layers progressively reduce the spatial or temporal resolution while increasing the channel dimension.
However, the initial layers of the encoder and the final layers of the decoder operate at the original resolution, leading to significant memory and computational requirements~\cite{beyer2024vitspeed}.
The wavelet packet transform (WPT), shown in Figure \ref{fig:wpt}, is a linear and invertible transform that performs an analogous operation. In each level of the WPT, the signal is divided into high- and low-frequency components, then downsampled by a factor of two. By recursively applying this process, the WPT allows spatial and temporal resolution to be traded off for frequency resolution with minimal computation and no loss of information. 
In WaLLoC, we exploit this property by sandwiching the learnable analysis and synthesis transforms between the WPT and its inverse---allowing all neural network layers to operate at low resolution.

(b) Loss of quality in lightweight encoders. Previous efforts use reduced hidden dimension and distillation to reduce the computational cost of pixel-based autoencoders but incur a significant loss of detail in the process~\cite{ollin2023taesd}. However, the WPT's ability to isolate important signal components from redundancies alleviates this issue. Additionally, it is possible to exploit asymmetry between the encoder and decoder. The decoder objective---disentangling mixed signal components---is difficult and requires a complex DNN-based transform. In contrast, the encoder objective---discarding signal redundancies---becomes trivial after applying the WPT. Thus, WaLLoC sandwiches an asymmetric autoencoder---consisting of a shallow, linear analysis transform and a deep, nonlinear synthesis transform---between the WPT and its inverse.

\vspace{-2mm}
\subsection{Achieving high compression ratio.} 
\vspace{-2mm}
Quantization is the primary mechanism used in lossy compression to reduce bit rate and achieve a high compression ratio. However, the GADR-AEs that provide good dimensionality reduction are not compatible with quantization. For example, quantization of Stable Diffusion's VAE latents leads to severe distortion~\cite{ollin2023taesd}
However if quantization is applied, very high compression ratios can be achieved via entropy coding. In WaLLoC, we incorporate an entropy bottleneck---additive noise applied during training that guarantees quantization resilience during inference~\cite{balle2017end}.
We optimize the noise scale for 8-bit quantization, allowing us to use standard lossless codecs (e.g PNG or WeBP) for entropy coding. This combination provides an additional compression multiplier of up to $12\times$ compared to reducing the dimension only. 

\vspace{-2mm}
\subsection{Achieving uniform dimensionality reduction.} 
\vspace{-2mm}
In addition to quantization, neural codecs achieve high compression ratios via a loss term that encourages sparse, rather than low-dimensional latents~\cite{balle2017end}. Using this objective, it is possible to drive the energy of many of the latent dimensions to zero~\cite{he2022elic}. However, this type of non-uniform dimensionality reduction is difficult to exploit in compressed-domain learning. 
In WaLLoC, the analysis transform uniformly reduces the dimension by a fixed rate, making it a suitable replacement for resolution reduction in accelerating downstream models.

\vspace{-2mm}
\subsection{WaLLoC Implementation}
\vspace{-2mm}
WaLLoC's encoder consists of five stages as shown in Figure~\ref{fig:walloc}: (1) wavelet packet transform (WPT) to trade-off spatial or temporal resolution with channel resolution (2) learned analysis transform to reduce dimensionality (3) companding to whiten the latent distribution (4) entropy bottleneck to provide resilience to quantization and (5) entropy coding to provide high compression ratios. The decoder consists of the reverse operations: (5) entropy decoding, (6) decompanding, (7) learned synthesis transform, and (8) inverse WPT. We now provide detailed explanations for each component.

\vspace{-5mm}
\paragraph{Wavelet packet transform.}
Figure \ref{fig:wpt} shows the workflow of the wavelet packet transform (WPT) and its inverse.
We use the Cohen–Daubechies–Feauveau (CDF) 9/7 wavelet~\cite{mallat2008} to construct the a dyadic filterbank consisting of highpass analysis ($\text{H}_{\text{A}}$),  lowpass analysis ($\text{L}_{\text{A}}$), highpass synthesis ($\text{H}_{\text{S}}$), and lowpass synthesis ($\text{L}_{\text{S}}$) filters.
The CDF 9/7 wavelet is chosen for its balance between computational efficiency and energy compaction.
Since these same filters are used in the JPEG 2000 standard, they are widely supported in software.
The WPT reduces the input resolution $R_{\textbf{x}}$ and increases the input channel count $C_{\textbf{x}}$ by a factor $2^J$ for 1D signals (audio) and by $4^J$ for 2D signals (images), but is linear and invertible. For stereo audio, we use $J=8$, resulting in $C_{\wstilde{\textbf{X}}}=512$ channels after the WPT. For RGB images, we use $J=3$, resulting in $C_{\wstilde{\textbf{X}}}=192$.

\vspace{-5mm}
\paragraph{Autoencoder and entropy bottleneck.}
The output of the WPT $\wtilde{\textbf{X}}$ is projected to a latent representation $\textbf{z}$ via a learnable analysis transform $\mathcal{G}_{\text{A}}$, which consists of a single linear layer. The latent dimension $C_\textbf{z}$ is a hyperparameter chosen based on the desired degree of dimensionality reduction. To achieve quantization-resilient latent representations, we adopt the entropy bottleneck method from end-to-end learned compression~\cite{balle2017end}, which consists of adding uniform noise $\mathcal{U}[-0.5,0.5]$ to the latent representation during training. Since the sub-band wavelet coefficients of many natural signals follow a generalized Gaussian distribution (GGD)~\cite{westerink1991subband}, we apply the Gaussian CDF $\Phi(\textbf{z})$ as a companding operation prior to the entropy bottleneck. Thus, the final encoder output is
$\hat{\textbf{z}}_{\text{t}}=\Phi(z)+\mathcal{U}$ during training and 
$\hat{\textbf{z}}_{\text{c}}=\text{round}\left(\Phi(\textbf{z})\right)$ during the compression pipeline. We scale the inputs and outputs of the companding operation $\Phi$ to guarantee latents in the range [-127, 127], which in turn guarantees that $\hat{\text{z}}_{\text{c}}$ does not underflow or overflow when quantized to a signed 8-bit integer.
The decoder consists of a learnable synthesis transform $\mathcal{G}_{\text{S}}$ followed by the IWPT. $\mathcal{G}_{\text{S}}$ is a convolutional neural network consisting the same residual blocks used in Stable Audio~\cite{evans2024stable} and Stable Diffusion~3~\cite{esser2024scaling} for 1D and 2D signals respectively.
We use a hidden dimension of $C_{\text{hidden}}=768$ for both the RGB image decoder and stereo audio decoders.
Additional implementation details are available in our public code repositories~\footnote{\href{https://github.com/danjacobellis/walloc/}{Code repository for WaLLoC}. \href{https://github.com/danjacobellis/walloc/}{Code and and experiments for compressed-domain learning.}}.

\vspace{-5mm}
\paragraph{Entropy coding.}
After quantization, an additional lossless compression step can be applied. We performed preliminary tests using zlib
, PNG (Deflate), and the lossless mode of WebP
. We found that WebP's entropy coding provided the best compression ratio---even for audio signals---while maintaining high throughput and compatibility with ML frameworks like PyTorch. Since WebP expects 24-bit RGB inputs, we rearrange the multi-channel 8-bit latent tensor into groups of three and concatenate channel groups along the temporal or spatial dimensions.

\vspace{-5mm}
\paragraph{Training.}
We train four codecs---two for stereo audio (5$\times$, 20$\times$) and two for RGB images (4$\times$, 20$\times$)---on The lossless MUSDB18-HQ~\cite{rafii2017musdb18} and LSDIR~\cite{li2023lsdir} datasets. In each case, the training objective is to minimize mean squared reconstruction error when latents are subjected to uniform additive noise in the range [-0.5,0.5]. 

\vspace{-3mm}
\section{Evaluation}
\vspace{-3mm}

We conduct a comprehensive evaluation of WaLLoC to demonstrate its efficacy for compressed domain learning. Our evaluation consists of two main parts. (1) \textbf{Compression trade-off analysis.} We compare WaLLoC against other lossy codecs in terms of the trade-off between dimensionality reduction, compression ratio, distortion, perceptual quality, and computation (Section~\ref{sec:compressionanalysis}). (2) \textbf{Compressed learning and resolution scaling.} We train and evaluate various machine learning models on representations produced by WaLLoC, and compare their resolution scaling properties to pixel-based and sample-based versions (Section~\ref{sec:resolutionscaling}).

\vspace{-2mm}
\subsection{Compression trade-off analysis}\label{sec:compressionanalysis}
\vspace{-2mm}
We compare WaLLoC against other popular conventional and neural codecs~\cite{esser2024scaling,evans2024stable,defossez2022high,cheng2020learned} across five key metrics: (1) degree of dimensionality reduction, (2) compression ratio, (3) distortion, (4) perceptual quality, and (5) computation.
For images, distortion is measured via PSNR and MS-SSIM~\cite{wang2004image}, while perceptual quality is evaluated via LPIPS\cite{zhang2018unreasonable} and DISTS~\cite{ding2020image}. For audio, distortion is measured via PSNR, SSDR, and SRDR~\cite{watcharasupat2024quantifying}, and perceptual quality is evaluated via CDPAM~\cite{manocha2021cdpam}. For both audio and images, the computational cost is measured in terms of average encoding and decoding throughput (megapixels or megasamples per second). Measurements are made on three different platforms: Low-power CPU (Raspberry Pi), High-power CPU (Intel i9), and GPU (RTX 4090).

\begin{table}[t]
\centering
\begin{tabular}{lrrrrrrrrrrrr}
\toprule
Method & DR & CR & Enc. & Dec. & PSNR & MS-SSIM & LPIPS$_{\text{dB}}$ & DISTS$_{\text{dB}}$ \\
\midrule
WEBP & 1 & \textbf{40.6} & \textbf{22.1} & \textbf{2746} & 28.2 & 0.96 & 5.94 & 13.1 \\
Cheng2020 & 4 & 21.8 & 0.289 & 0.139 & \textbf{33.8} & \textbf{0.99} & \underline{8.82} & \underline{16.9} \\
\rowcolor{lightgray}
WaLLoC & 4 & 8.53 & \underline{14.0} & \underline{0.47} & \underline{33.5} & \textbf{0.99} & \textbf{11.2} & \textbf{19.3} \\
SD 3.0 & \underline{12} & 6.00 & 0.195 & 0.101 & 20.9 & 0.84 & 8.33 & 13.8 \\
\rowcolor{lightgray}
WaLLoC & \textbf{16} & \underline{35.2} & \textbf{22.1} & 0.466 & 27.5 & \underline{0.97} & 6.51 & 13.9 \\
\bottomrule
\end{tabular}
\vspace{-4mm}
\caption{RGB image compression comparison. Metrics: dimensionality reduction (DR), compression ratio (CR), encoding (Enc.) and decoding (Dec.) throughput (Megapixels/sec, CPU), distortion (PSNR, MS-SSIM) and perceptual quality (LPIPS$_{\text{dB}}$, DISTS$_{\text{dB}}$). We report LPIPS$_{\text{dB}}=-10\log_{10}(\text{LPIPS})$ and DISTS$_{\text{dB}}=-10\log_{10}(\text{DISTS})$ so that higher values are better for each metric. For each metric, the best performing method is in boldface and the second best is underlined.}
\label{tab:RGB}
\end{table}

\vspace{-5mm}
\paragraph{Results of compression trade-off analysis.}
Figure \ref{fig:radar}, Table~\ref{tab:RGB}, and Table \ref{tab:stereo} summarize the trade-offs between rate, distortion, perception, computation, and dimension between different types of compression.
For RGB Images, WaLLoC achieves nearly 12$\times$ higher compression ratio (35:1 vs 6:1) compared to the VAE used in Stable Diffusion 3, despite offering a higher degree of dimensionality reduction (16$\times$ vs 12$\times$) and similar quality (13.9 dB vs 13.8 dB DISTS).
Compared to Cheng et al.~\cite{cheng2020learned}, WaLLoC achieves more than 48$\times$ higher encoding throughput (14.0 vs 0.29 MPix/sec) and similar quality (19.3 dB vs 16.9 dB DISTS).
For stereo audio, WaLLoC achieves significantly higher spatial quality (22.5 dB vs 15.7 dB SSDR) than Stable Audio's VAE, but with more than $300\times$ higher encoding throughput. Examples of decoded images from the LSDIR validation set are provided on Hugging Face
\footnote{\href{https://huggingface.co/datasets/danjacobellis/LSDIR_RGB_Li_48c_J3_nf8}{Examples of decoded images}}. Additional results, including GPU and Raspberry Pi throughput, are available in our code repository
\footnote{\href{https://github.com/danjacobellis/LCCL}{Repository containing full code, experiments, and results}}.

\begin{table}[t]
\centering
\begin{tabular}{lrrrrrrrr}
\toprule
Method & DR & CR & Enc & Dec & PSNR & SSDR & SRDR & CDPAM \\
\midrule
Opus & 1.0 & \textbf{119} & 11.5 & \textbf{102} & 30.4 & 16.7 & 5.03 & 40.4 \\
\rowcolor{lightgray}
WaLLoC & 4.74 & 21.3 & \underline{77.8} & 11.2 & \textbf{39.0} & \textbf{33.3} & \textbf{13.9} & 41.1 \\
EnCodec & 5.0 & \underline{114} & 2.75 & 3.03 & 31.9 & \underline{22.7} & 6.69 & \underline{47.4} \\
\rowcolor{lightgray}
WaLLoC & \underline{18.9} & 76.3 & \textbf{121} & \underline{12.2} & \underline{33.3} & 22.5 & \underline{8.06} & 36.6 \\
Stable Audio & \textbf{64.0} & 64.0 & 0.308 & 0.30 & 28.4 & 15.7 & 2.03 & \textbf{49.7} \\
\bottomrule
\end{tabular}
\vspace{-8mm}
\caption{Stereo audio compression results. Abbreviations are the same as Table \ref{tab:RGB}.
}

\label{tab:stereo}
\end{table}

\vspace{-2mm}
\subsection{Compressed learning and resolution scaling}\label{sec:resolutionscaling}
\vspace{-1mm}

Next, we describe our methodology for evaluating compressed domain learning.  

(a) Applications, models, and datasets. We evaluate WaLLoC on 4 machine perception tasks: (1) image classification, (2) image colorization, (3) document understanding and (4) music source separation. For classification and colorization, we train ViT-Ti models
with conditional position encoding~\cite{tu2022maxvit} on the ImageNet-1k dataset.
For music source separation, we train a CNN to separate the vocal track from music segments in MUSDB18-HQ. The CNN consists of 12 identical convolutional layers structured identically to Stable Audio's mid block~\cite{evans2024stable}.
For document understanding, we use PaliGemma~\cite{beyer2024paligemma}  fine-tuned at varying resolution on the DocVQA~\cite{mathew2021docvqa} dataset, and report the average normalized levenshtein similarity (ANLS) on the test set.

(b) Resolution scaling strategy. For \textbf{image classification}, we reduce the input sequence length by 4$\times$ or 16$\times$ compared to the baseline of $256^2$ pixels and $16^2$ patches, but keep the area of each patch constant ($1/16^2$). We report the accuracy of models trained on reduced resolution inputs with models trained on the identically WaLLoC latents. For \textbf{document understanding}, training models on the scale of PaliGemma is outside the scope of this work. Instead, we evaluate on decoded WaLLoC representations using the highest-resolution PaliGemma variant ($896^2$). To emulate the effect of resolution reduction with this high-resolution variant, we downsample images to the desired resolution, ($224^2$ or $448^2$), then apply Lanczos resampling to interpolate back to $896^2$. For \textbf{Image colorization} and \textbf{music source separation}, we increase the input patch size proportionally to the resolution to keep the sequence length---and therefore the required computation---roughly constant.

\paragraph{Results of compressed-domain learning and resolution scaling.}
Figure~\ref{fig:resolution} shows the improvement in performance when using WaLLoC-derived representations instead of resolution reduction. Across each of the four tasks, WaLLoC provides superior accuracy to naive resolution reduction while providing the same improvement in latency and memory consumption. For discriminative models, WalloC profoundly increases accuracy of efficient image classification (50.6\% vs 23.1\% accuracy) and  document understanding (81.1 vs 43.7 ANLS). For signal enhancement, WaLLoC provides superior scaling to high resolution and large patches---offering a 16.7\% improvement in colorization LPIPS and a 3.1 dB improvement in PSNR for source separation.

\section{Conclusion and future work}
We introduced WaLLoC, a compression framework to support compressed-domain learning.
Our experiments demonstrate that WaLLoC significantly accelerates downstream models without sacrificing accuracy, achieving up to 20× dimensionality reduction with minimal encoding cost.
Future work will explore extending WaLLoC to applications involving high-resolution signal types for which existing compression methods fall short, such as hyperspectral images or whole-slide microscopy.
These domains present additional challenges but also offer greater potential benefits due to increased signal redundancies.

\section{Acknowledgments}
\vspace{-3mm}
We thank the anonymous reviewers for their helpful feedback. We thank the members of the UT-SysML research group for their insightful discussions to improve this work. This work was supported by the UT ECE junior faculty start-up fund, UT iMAGiNE consortium and its industrial affiliates, an award from the UT Machine Learning Lab (MLL), the AMD Chair Endowment, the Cisco Research Award, and the Amazon Research Award.

\FloatBarrier

\section{References}
\vspace{-2mm}
\bibliographystyle{IEEEbib}
\bibliography{refs}

\clearpage
\Section{Appendix}
\vspace{-7mm}
\begin{figure}[!h]
    \centering
    \includegraphics[width=\textwidth]{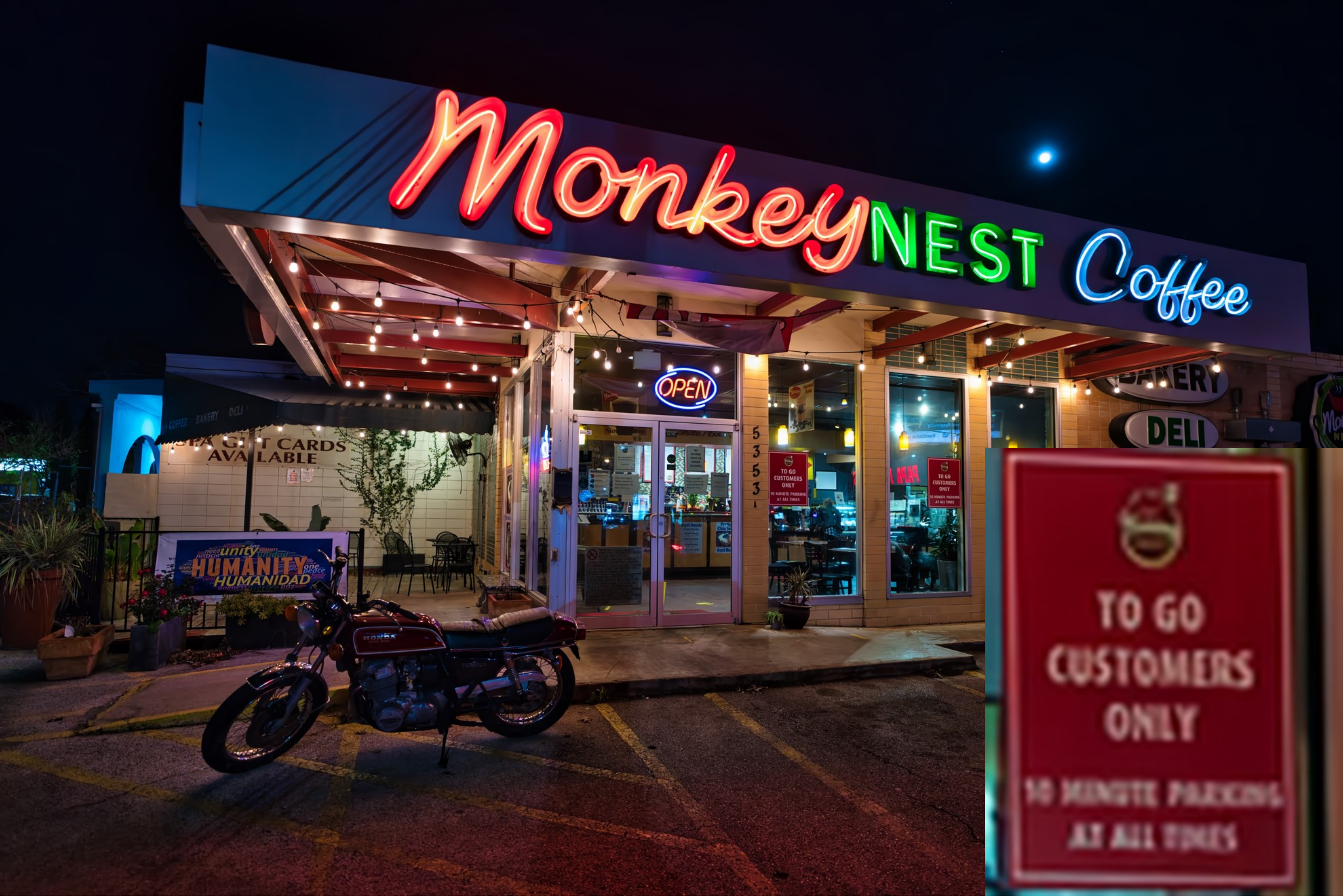}
    \vspace{-13mm}
    \caption{Cheng et al. 2020 \cite{cheng2020learned}}
    \label{fig:cheng2020}
    \vspace{-11mm}
\end{figure}

\begin{figure}[!h]
    \centering
    \includegraphics[width=\textwidth]{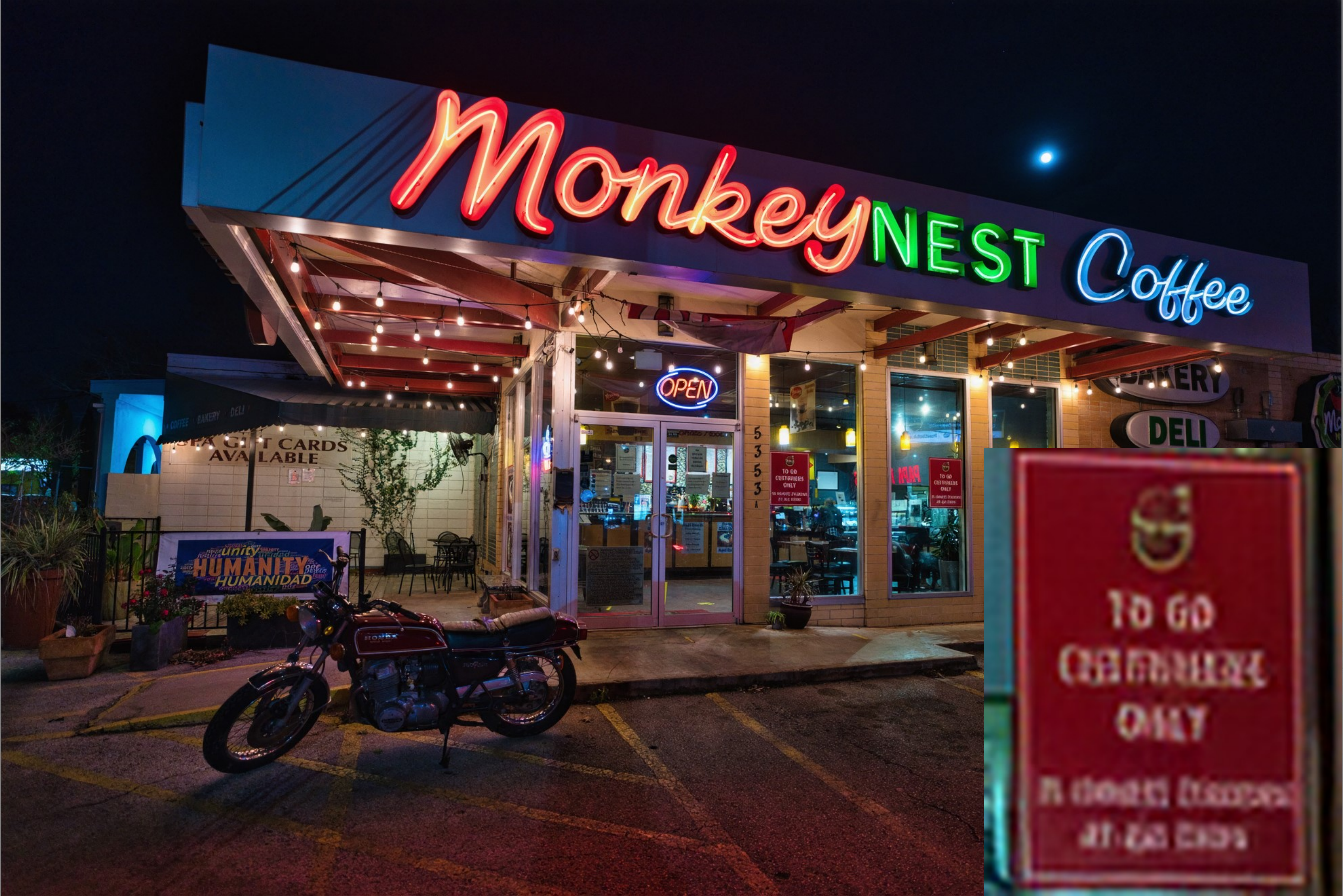}
    \vspace{-13mm}
    \caption{Stable Diffusion 3 VAE \cite{esser2024scaling}}
    \label{fig:sd3}
    \vspace{-10mm}
\end{figure}

\begin{figure}[!h]
    \centering
    \includegraphics[width=\textwidth]{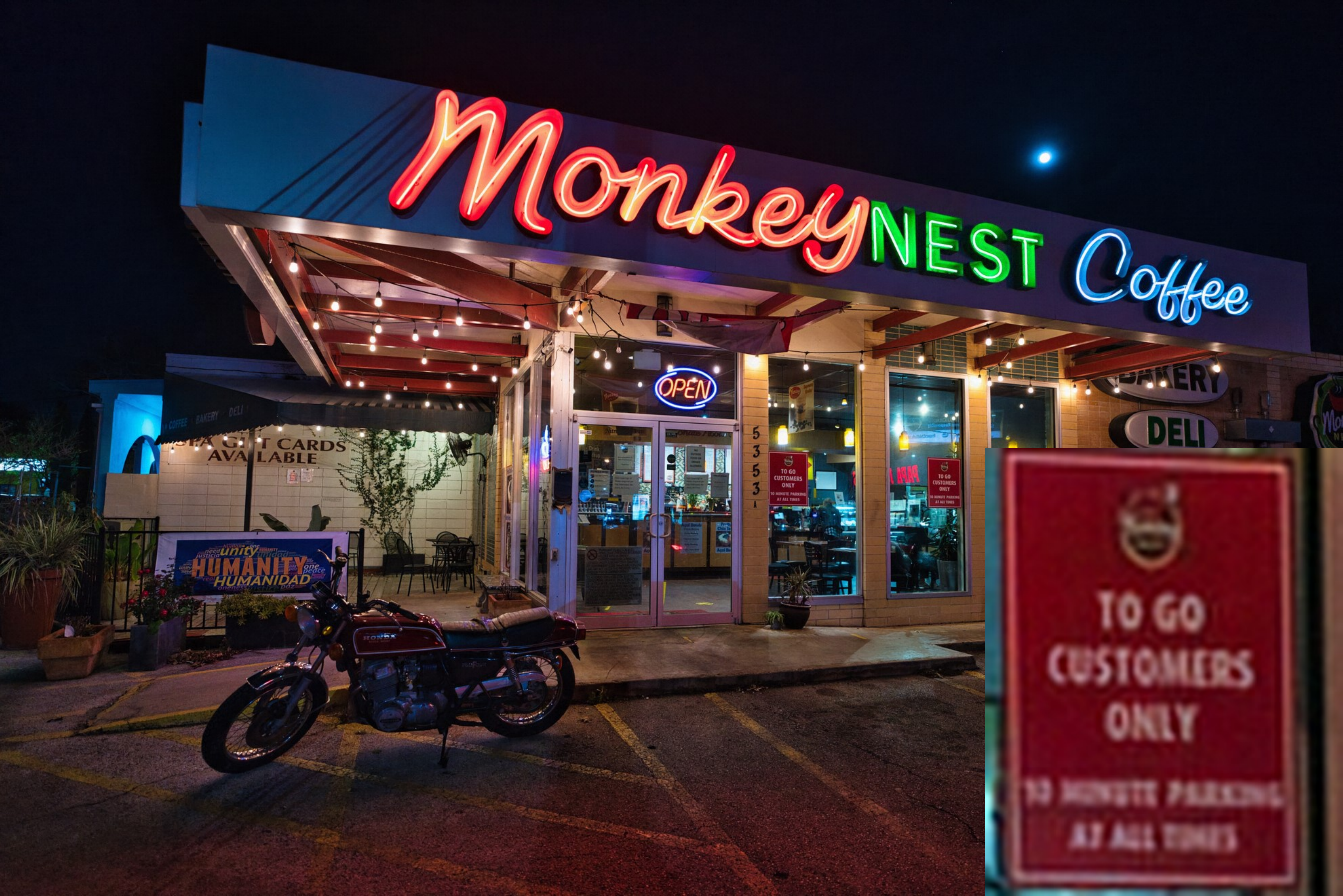}
    \vspace{-13mm}
    \caption{WaLLoC 4$\times$}
    \label{fig:walloc4x}
\end{figure}

\begin{figure}[!h]
    \centering
    \includegraphics[width=\textwidth]{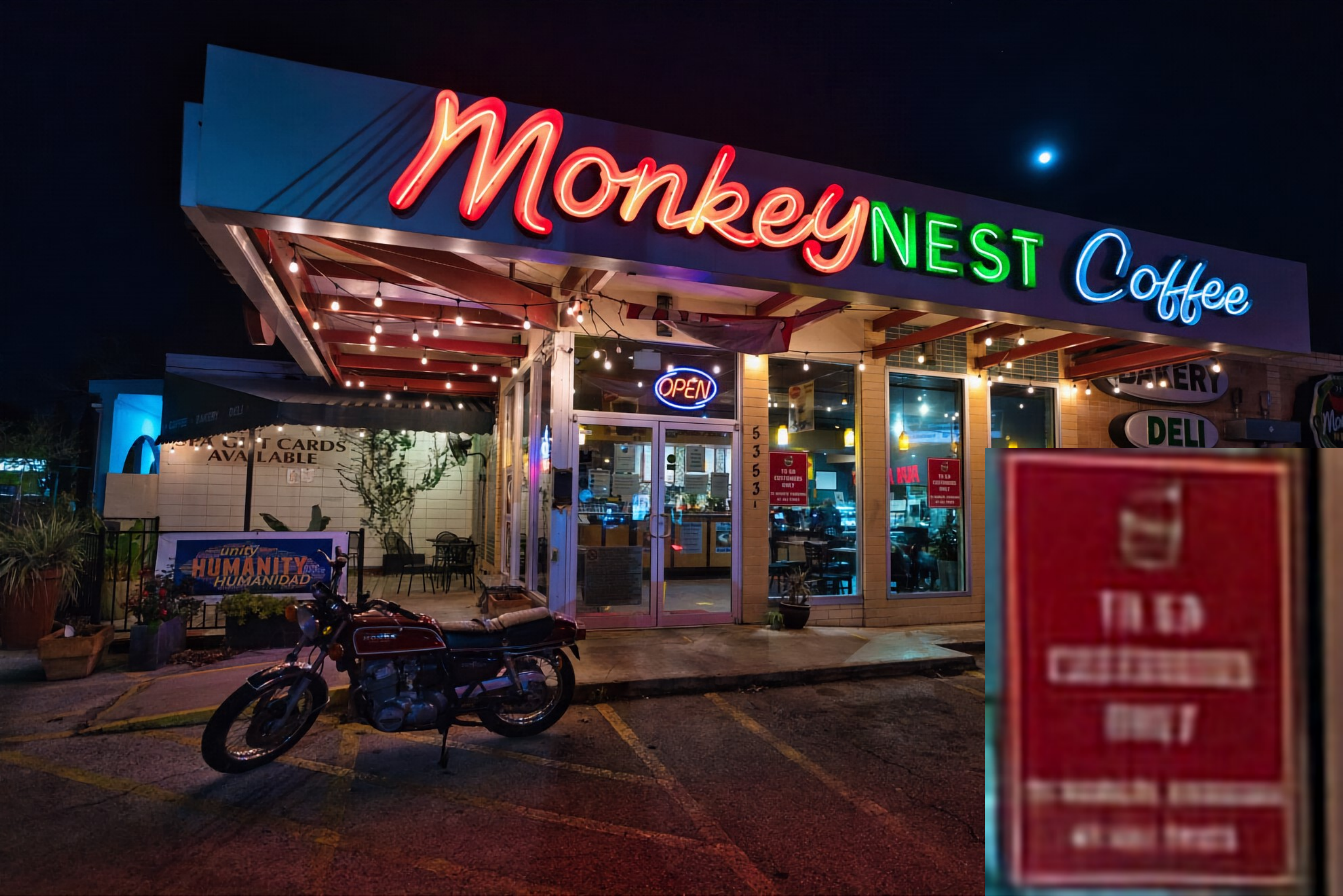}
    \vspace{-13mm}
    \caption{WaLLoC 16$\times$}
    \label{fig:walloc16x}
    \vspace{-10mm}
\end{figure}

\clearpage

\begin{figure}[!h]
    \centering
    \includegraphics[width=\textwidth]{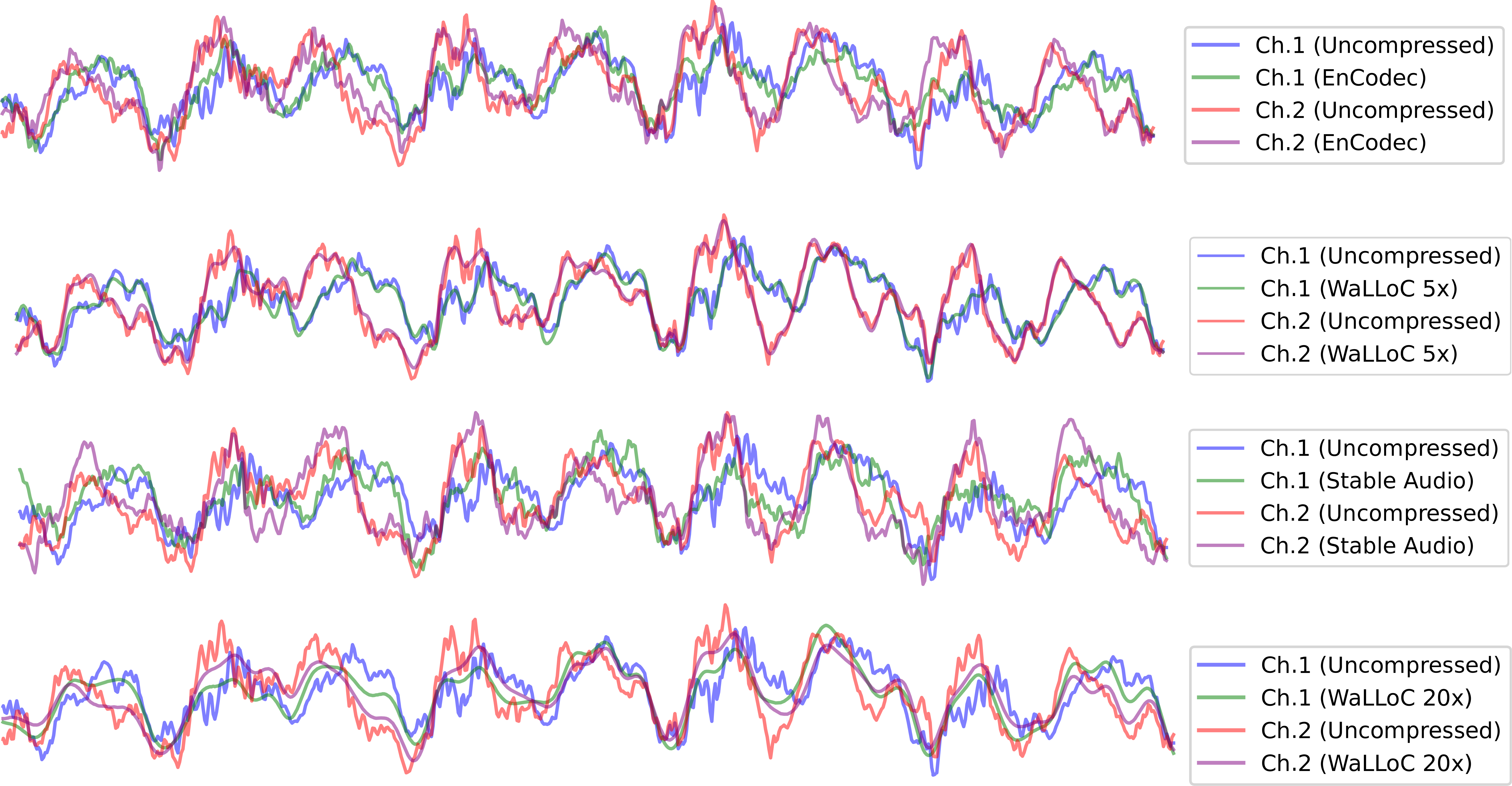}
    \caption{Stereo reconstruction of an audio segment from the MUSDB test set.}
    \label{fig:audiocomparison}
\end{figure}

\clearpage

\begin{figure}[!h]
    \centering
    \includegraphics[width=\textwidth]{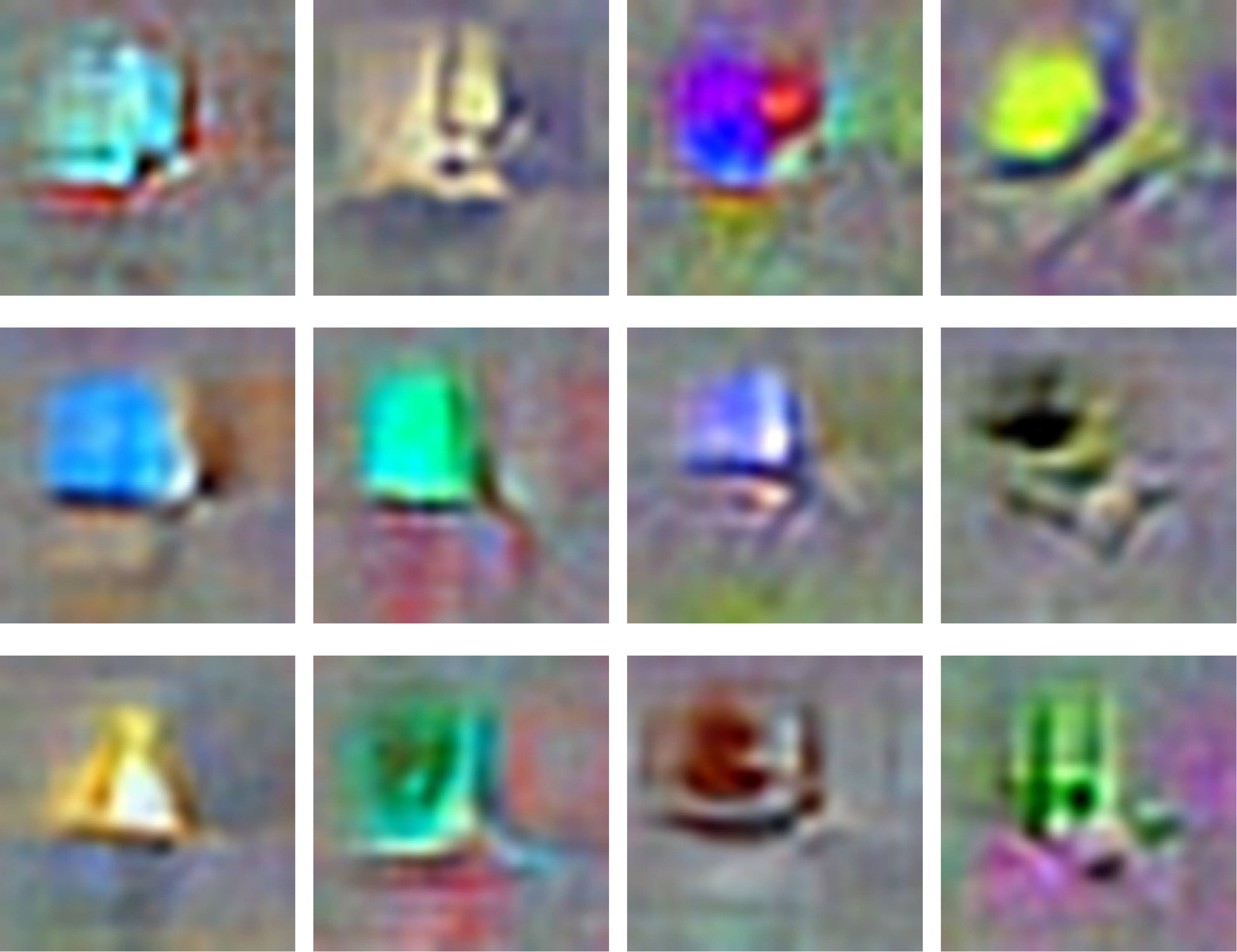}
    \caption{Result of using the $C_{\textbf{z}}=12$ RGB codec (WaLLoC 16$\times$) to decode a \(12 \times 3 \times 3\) latent with all elements equal to zero except except for channel $i$, which is set to $\begin{bmatrix} 0 & 0 & 0 \\ 0 & 31 & 0 \\ 0 & 0 & 0 \end{bmatrix}$.
}
    \label{fig:12c}
\end{figure}

\begin{figure}[!h]
    \centering
    \includegraphics[width=\textwidth]{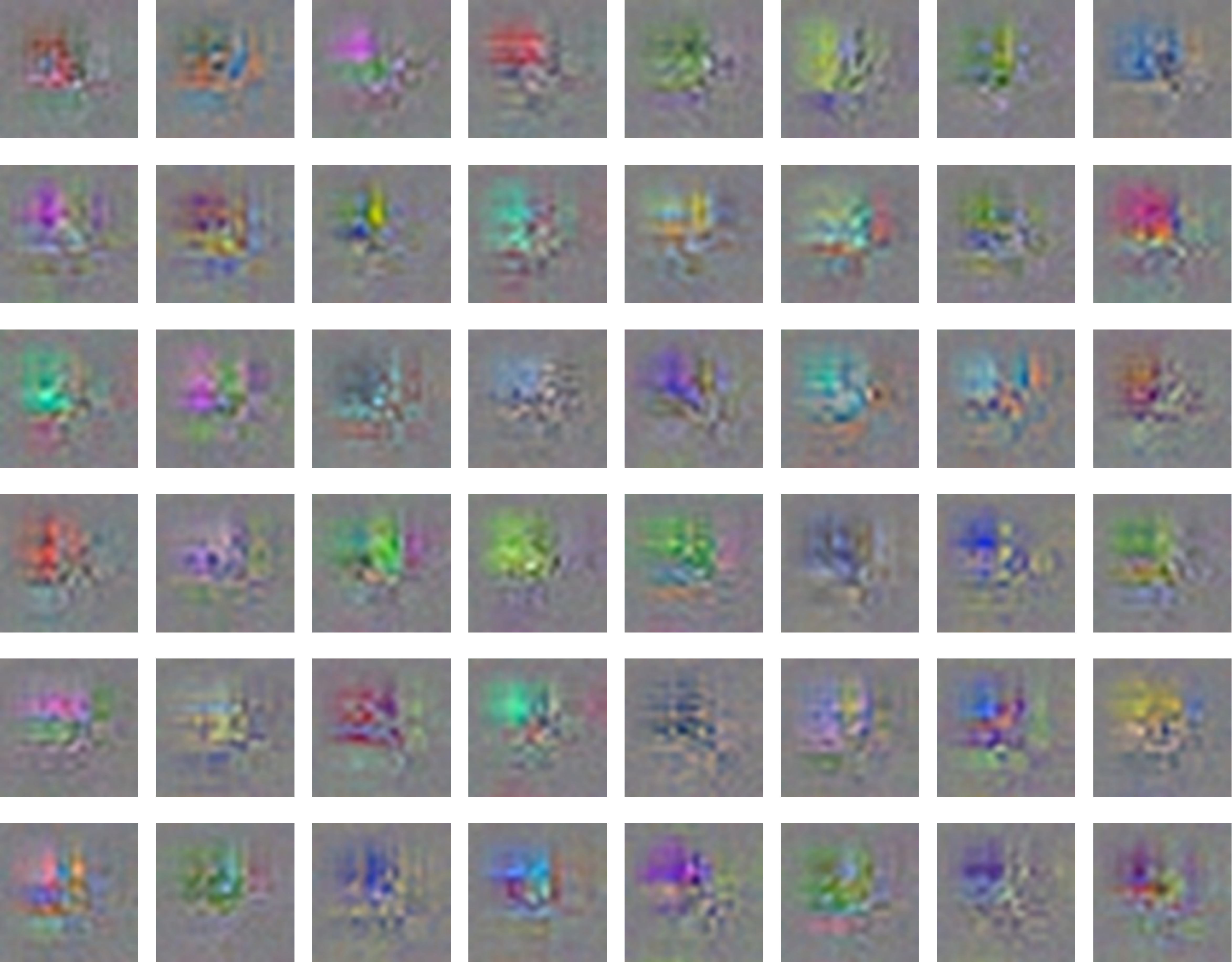}
    \caption{Result of using the $C_{\textbf{z}}=48$ RGB codec (WaLLoC 4$\times$) to decode a \(48 \times 3 \times 3\) latent with all elements equal to zero except except for channel $i$, which is set to $\begin{bmatrix} 0 & 0 & 0 \\ 0 & 31 & 0 \\ 0 & 0 & 0 \end{bmatrix}$.
}
    \label{fig:48c}
\end{figure}

\begin{table}[!h]
\centering
\begin{tabular}{@{}cccccc@{}}
\toprule
\textbf{Task} & \textbf{\shortstack{Resolution\\Equivalent}} & \textbf{\shortstack{WaLLoC\\Variant}} & \textbf{\shortstack{Performance\\(Resize)}} & \textbf{\shortstack{Performance\\(Compress)}} & \textbf{Change} \\ \midrule
\multirow{3}{*}{\shortstack{Classification\\(Acc., \%)}}  
& $64^2$ px  & 16$\times$ & 23.1  & 50.3  & $\uparrow$27.2 \\
& $128^2$ px & 4$\times$  & 55.8  & 64.3  & $\uparrow$8.5  \\
& $256^2$ px & --          & 71.1  & --    & --             \\ \midrule
\multirow{3}{*}{\shortstack{Doc. VQA\\(ANLS)}} 
& $224^2$ px & 16$\times$  & 43.7  & 81.1  & $\uparrow$37.4 \\
& $448^2$ px & 4$\times$   & 78.0  & 84.1  & $\uparrow$6.1  \\
& $896^2$ px & --          & 84.8  & --    & --             \\ \midrule
\multirow{3}{*}{\shortstack{Colorization\\(LPIPS, dB)}}  
& $128^2$ px & --          & 1.76  & --    & --             \\
& $256^2$ px & 4$\times$   & 2.33  & 2.47  & $\uparrow$0.14          \\
& $512^2$ px & 16$\times$  & 2.43  & 2.83  & $\uparrow$0.40          \\ \midrule
\multirow{3}{*}{\shortstack{Source sep.\\(PSNR, dB)}} 
& 2.4 kHz    & --          & 31.1  & --    & --             \\
& 11 kHz     & 5$\times$   & 32.0  & 34.4  & $\uparrow$2.4  \\
& 44 kHz     & 18$\times$  & 31.8  & 34.2  & $\uparrow$2.4  \\ \bottomrule
\end{tabular}
\caption{Results of resolution scaling experiments.}
\end{table}

\end{document}